# Multibeam Satellite Frequency/Time Duality Study and Capacity Optimization

Jiang Lei, *Student Member, IEEE,* M. A. Vázquez-Castro, *Senior Member, IEEE,*

***Abstract*—In this paper, we investigate two new candidate transmission schemes, Non-Orthogonal Frequency Reuse (NOFR) and Beam-Hoping (BH). They operate in different domains (frequency and time/space, respectively), and we want to know which domain shows overall best performance. We propose a novel formulation of the Signal-to-Interference plus Noise Ratio (SINR) which allows us to prove the frequency/time duality of these schemes. Further, we propose two novel capacity optimization approaches assuming per-beam SINR constraints in order to use the satellite resources (e.g. power and bandwidth) more efficiently. Moreover, we develop a general methodology to include technological constraints due to realistic implementations, and obtain the main factors that prevent the two technologies dual of each other in practice, and formulate the technological gap between them. The Shannon capacity (upper bound) and current state-of-the-art coding and modulations are analyzed in order to quantify the gap and to evaluate the performance of the two candidate schemes. Simulation results show significant improvements in terms of power gain, spectral efficiency and traffic matching ratio when comparing with conventional systems, which are designed based on uniform bandwidth and power allocation. The results also show that BH system turns out to show a less complex design and performs better than NOFR system specially for non-real time services.**

## I. Introduction

THE efficient management of multibeam antenna resources, e.g. power, bandwidth and time-slot, is crucial for economic competitiveness. Specifically, in modern satellite networks, each satellite uses multiple beams, each of which illuminates a cell on the ground to serve a coverage area. Multibeam antenna technology is used because it can increase the total system capacity significantly [1]–[3]. However, each beam will compete with others for resources to achieve satisfactory communication. This is due to the fact that the traffic demand throughout the coverage is potentially highly asymmetrical among the beams. Therefore, the satellite requires a certain degree of flexibility in allocating the power, bandwidth and time-slot resources to achieve a good match between offered and requested traffic.

There are some precedents of resource allocation optimization techniques for satellite systems. In [4], [5], the authors investigate the dynamic bandwidth allocation techniques, and in [6] the authors propose a quasi-optimal solution to manage the frequency slots allocation to service providers, however, the results are only for the satellite uplink. A power allocation policy is proposed in [7], which suggests to stabilize the system based on the amount of unfinished work in the queue and the channel state, and a routing decision is made for the maximum total throughput. However, the authors do not taking into account the co-channel interference. In [8], [9], a tradeoff strategy is proposed between different objectives and system optimization. The power allocation is optimized based on the traffic distribution and channel conditions. However, the co-channel interference is not taken into account. In [10], a joint power and carrier allocation problem is discussed, however, it focus on the return (RTN) uplink. In [11], [12], the authors focused on the capacity optimization in multibeam satellite system, and the duality in frequency and time domain is studied. The optimization problem of power and carrier allocation has also been addressed in terrestrial networks. E.g., the authors in [13] propose an axiomatic-based interference model for Signal-to-Interference plus Noise (SINR) balancing problem, but the conclusions cannot be directly extrapolated to a satellite scenario. Although the resource allocation optimization has been study extensively, the objective of this paper is different from the aforementioned literatures in various aspects. E.g., most of the existing resource allocation optimization approaches focus on terrestrial networks or on the satellite RTN link, while we focus on the satellite FWD link. In addition, we address the resource allocation according to the realistic asymmetric traffic distribution by *managing the co-channel interference* due to frequency reuse. Existing work has focused on the analysis of the resources allocation in frequency domain. Our aim is to characterize the best resource allocation scheme in multi-domain, and to show in which domain the overall performance is best.

In this paper, we investigate two new transmission schemes, Non-Orthogonal Frequency Reuse (NOFR) technique and Beam-Hopping (BH), which have been chosen as candidates to replace current regular frequency reuse transmission scheme. The first one is designed based on the frequency division over a flexible payload design which allows managing interference as an alternative to a complete orthogonal frequency reuse. The second one is based on the time/space division. Both techniques can potentially cope with the asymmetric traffic distribution as opposed to current satellite resources allocation scheme, which is designed to allocate fixed power and bandwidth to each ground cell. This leads to a waste of resources in low traffic requirement beams. On the contrary, it does not satisfy traffic demand in the hot ground cells, where the traffic requirement is high.

In this paper, we study the duality between NOFR and BH, i.e. frequency and time/space duality. The concept of

Jiang Lei and M. A. Vázquez-Castro are with the Department of Telecommunications and System Engineering, Universitat Autónoma de Barcelona, Barcelona 08193, Spain (e-mail: jiang.lei@uab.cat; angeles.vazquez@uab.cat).

duality gives rise to many interesting properties to simplify system models. Generally speaking, a duality translates concepts, theorems or mathematical structures into other concepts, theorems or structures, in a one-to-one fashion, often (but not always) by means of an involution operation: if the dual of $A$ is $B$, then the dual of $B$ is $A$. E.g., the duality of space and time is studied in [14], [15], Gaussian multiple-access/broadcast channel duality is discussed in [16], uplink and downlink duality is presented in [17], [18]. We develop a general methodology to study the duality of the two schemes that also considers the technological constraints due to realistic implementations, and obtain the main factors that prevent the two schemes be in practice dual of each other.

The novel contributions of this paper can be summarized as follows:

- The frequency and time duality is formulated for the multibeam satellite system, and the duality conditions are derived for a practical system.
- We prove that new transmission schemes, NOFR and BH, can match much better than the conventional design in the realistic asymmetric traffic model, and also prove that the BH system performs only slightly better than NOFR.

The rest of this paper is organized as follows: In Section II, the problem statement is presented. In Section III, we model the multibeam downlink system to obtain a mathematical expression of SINR. The duality of NOFR and BH is discussed in Section IV. In Section V, we formulate and solve the satellite capacity optimization problems. In Section VI, the technological gap is obtained with a realistic system payload model. The simulation results are presented in Section VII. In Section VIII, we draw the conclusions of the paper.

## II. Problem Statement

In multibeam satellite systems, the beamforming antenna generates $K$ beams over the coverage area. For both NOFR and BH systems, we firstly introduce some payload parameters (as shown in Table I).

- Granularity: $B_c$ is the carrier granularity defined as $B_c = B_{\text{tot}}/N_c$ in NOFR system. It means that the allocated bandwidth per ground cell should be an integral multiple of $B_c$. We use $T_s$, with the same meaning but in BH system, i.e. the minimum unit of time duration that can be allocated per cell.
- Resource allocation matrix: $w_{ij}$ and $t_{ij}$ are the elements of the resource allocation matrix for the NOFR and BH systems, respectively. The matrix indicates which carrier or time-slot $j$ is allocated to the ground cell $i$. Note that BH can direct the satellite beams to specific ground cells, i.e. it is a space allocation too.

In the case of a NOFR scheme, each ground cell can be allocated a variable number of carriers (e.g. $N_i$, as shown in Fig.1) depending on the traffic requirement. Carriers can be re-used throughout the coverage, but we do not impose any restrictions on the frequency reuse, it will be given by the resource optimization (i.e. interference minimization for a given traffic demand pattern) and therefore will be non-orthogonal. In the case of a BH system, the total bandwidth is simultaneously used by a set of the ground cells during a time-slot ($T_s$). We assume that the resource allocation takes place during a given time window divided into $N_t$ time-slot. Each ground cell can be allocated a variable number of time-slot.

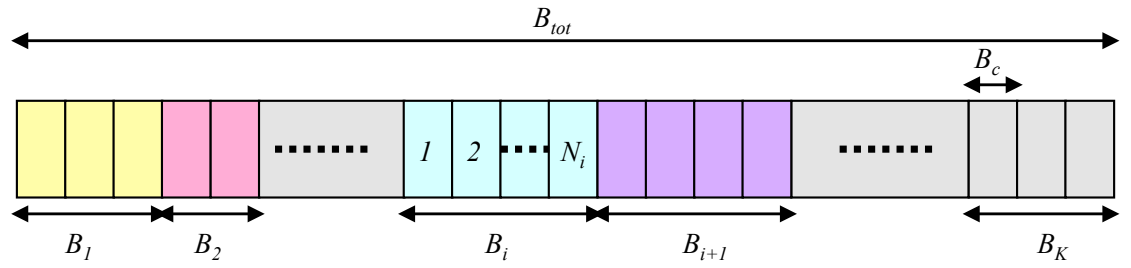

Fig. 1. Bandwidth segmentation.

Note that both techniques allow a number of ground cells to use the same frequency band or time-slot, resulting in co-channel interference. The problem tackled in this paper is to optimize the capacity by taking into account the co-channel interference. Further, we prove the duality of these techniques by developing a formulation that also allows including technological constraints. Moreover, we compare the performance of the proposed new transmission schemes with the current one for the realistic asymmetric traffic model. To do so, we propose a novel iterative algorithm, which do not only optimize the power and bandwidth allocation (for NOFR systems), but also optimize the structure of the spectral mask matrix $\mathbf{W}$. This matrix indicates which carriers are allocated per-beam in order to minimize the co-channel interference. Although the power and carrier optimization problem has been addressed in terrestrial networks, it is new in satellite communications.

## III. Multibeam System Model

In this section, we first formulate the multibeam system model in frequency domain (i.e. for the NOFR scheme). After that we state the conditions for duality and prove that NOFR and BH are dual of each other and hence, the formulation is also valid in time domain(i.e. for the BH scheme). This dual formulation allows us to derive a unique SINR expression, which will be used in the following section for capacity optimization. Following, we introduce the different sub-models.

### A. Channel Model

We do an analysis in time and hence the channel attenuation corresponds to the free space losses and atmospheric losses (in case of frequencies above Ka band). We assume an instantaneous analysis with fixed coefficient. The channel attenuation amplitude matrix $\mathbf{A} \in \mathbb{C}^{K\times K}$ is defined as

$$\mathbf{A} = \text{diag}\left\{\alpha_1, \alpha_2, \cdots, \alpha_K\right\}, \quad (1)$$

where $\alpha_i$ denotes the channel attenuation factor over the destination user beam $i$.

### B. Antenna Model

An Array Feed Reflector (AFR) based Antenna system is assumed, it can generate a regular beam grid array consisting of a very high number of highly overlapping, narrow beam width, composite user beams. Each beam is synthesized by adding array elements whose phases and amplitudes are adjustable, and hence we can provide flexible power allocation by controlling the On-Board Processor (OBP). Therefore, we suppose that the antenna gain matrix $\mathbf{G} \in \mathbb{C}^{K\times K}$ is given as

$$\mathbf{G} = \begin{bmatrix} g_{11} & g_{12} & \cdots & g_{1K} \\ g_{21} & g_{22} & \cdots & g_{2K} \\ \vdots & \vdots & \ddots & \vdots \\ g_{K1} & g_{K2} & \cdots & g_{KK} \end{bmatrix}, \quad (2)$$

where $|g_{ij}|^2 \in \mathbb{R}^{1\times 1}$ is the antenna gain of the on-board antenna feeds for the $j$th beam towards the $i$th user beam.

### C. Received Signal Model

In the frequency domain, the transmitted symbols over $N_c$ carriers to beam $i$ $(i = 1, 2, \cdots, K)$ are defined as $\mathbf{x}_i = [x_{i1}, x_{i2}, \cdots, x_{iN_c}]^T$. Let the spectral mask matrix $\mathbf{W} \in \mathbb{R}^{N_c\times K}$ be defined as $\mathbf{W} = [\mathbf{w}_1, \mathbf{w}_2, \cdots, \mathbf{w}_K]$, and the $i$th column vector $\mathbf{w}_i \in \mathbb{R}^{N_c\times 1}$ be defined as $\mathbf{w}_i = [w_{i1}, w_{i2}, \cdots, w_{iN_c}]^T$, which is the spectral mask vector for beam $i$ and indicates which TDM carriers and how much power is allocated to beam $i$.

Let $\mathbf{H} = \mathbf{AG}$ be the overall channel matrix, and $\mathbf{W}_i = \mathrm{diag}\{\mathbf{w}_i\}$. Then the received signal by all the $N_c$ carriers for beam $i$, $\mathbf{y}_i \in \mathbb{C}^{N_c\times 1}$, can be expressed as desired signal and interference as

$$\mathbf{y}_i = h_{ii}\tilde{\mathbf{x}}_i + \sum_{k=1(k\neq i)}^{K} h_{ik}\tilde{\mathbf{x}}_k + \mathbf{n}_i, \quad (3)$$

where $\tilde{\mathbf{x}}_i$ is the spectral masked symbol vector for beam $i$, defined as $\tilde{\mathbf{x}}_i = \mathbf{W}_i\mathbf{x}_i$. The first term corresponds to the desired signals coming from the $i$th on-board antenna. The second term is the sum of interference signals from the other on-board antennas. $\mathbf{n}_i \in \mathbb{C}^{N_c\times 1}$ is a column vector of zero-mean complex circular Gaussian noise with variance $\sigma^2$ at beam $i$.

### D. Signal-to-Interference plus Noise Ratio

In the frequency domain, the bandwidth is segmented as shown in Fig.1. We assume that the whole bandwidth is segmented in $N_c$ carriers. The spectral mask matrix can be reformulated as $\mathbf{W} = [\tilde{\mathbf{w}}_1^T, \tilde{\mathbf{w}}_2^T, \cdots, \tilde{\mathbf{w}}_{N_c}^T]^T$, where $\tilde{\mathbf{w}}_j = [w_{1j}, w_{2j}, \cdots, w_{Kj}]$, indicates which beams are allocated carrier $j$. Let the $i$th row of $\mathbf{H}$ be defined as $\mathbf{h}_i = [h_{i1}, h_{i2}, \cdots, h_{iK}]$ and $\tilde{\mathbf{h}}_i = \mathbf{h}_i|_{(h_{ii}=0)}$ is the channel of interference contribution. We assume that the amplitude of the transmitted symbols is normalized (i.e. $|x_{ij}|^2 = 1, \forall i = 1, \cdots, K; \forall j = 1, \cdots, N_c$).

Then, the transmitted signal power of all the carriers for beam $i$ can be given by the diagonal elements of the matrix $\mathbf{S}_i^f \in \mathbb{R}^{N_c\times N_c}$ as (note that the superscript $f$ and $t$ indicate the expression in frequency and time domain, respectively)

$$\mathbf{S}_i^f = |h_{ii}|^2\mathbf{W}_i\mathbf{W}_i^H. \quad (4)$$

And the co-channel interference power of all the carriers for beam $i$ can also be given by the diagonal elements of the matrix $\mathbf{U}_i^f \in \mathbb{R}^{N_c\times N_c}$ as

$$\mathbf{U}_i^f = \mathrm{diag}\left\{\left[\tilde{\mathbf{h}}_i\tilde{\mathbf{w}}_j^H\tilde{\mathbf{w}}_j\tilde{\mathbf{h}}_i^H\right]_{j=1,2,\cdots,N_c}\right\}. \quad (5)$$

TABLE I
FREQUENCY-TIME DUALITY

| | Frequency domain | Time domain |
|---|---|---|
| Granularity | $B_c$ | $T_s$ |
| Total number of carriers/time-slot | $N_c$ | $N_t$ |
| Resource allocation matrix | $w_{ij}$ | $t_{ij}$ |
| SINR ($\gamma_{ij}$) | $\gamma_{ij}^f$ | $\gamma_{ij}^t$ |
| Spectral efficiency ($\eta_{ij}$) | $\eta_{ij}^f$ | $\eta_{ij}^t$ |
| Throughput for beam $i$ | $R_i^f$ | $R_i^t$ |

Thus, the interference power plus the noise matrix, $\mathbf{V}_i^f$, will be given as

$$\mathbf{V}_i^f = \mathbf{U}_i^f + \sigma^2\mathbf{I}_{N_c}. \quad (6)$$

Consequently, the SINR for beam $i$, defined as $\mathbf{\Gamma}_i^f \in \mathbb{R}^{N_c\times N_c}$, can be expressed as

$$\mathbf{\Gamma}_i^f = \mathbf{S}_i^f(\mathbf{V}_i^f)^{-1}. \quad (7)$$

Obviously, $\mathbf{\Gamma}_i^f$ is a diagonal matrix, because both $\mathbf{S}_i^f$ and $\mathbf{V}_i^f$ are diagonal matrixes. Thus, the SINR for the $j$th carrier used by beam $i$ will be the $j$th diagonal element of the matrix $\mathbf{\Gamma}_i^f$. This means that for each carrier $j$ of beam $i$, the SINR can be formulated as

$$\gamma_{ij}^f = \frac{|h_{ii}w_{ij}|^2}{\sum_{k=1(k\neq i)}^{K} |h_{ik}w_{kj}|^2 + \sigma^2}. \quad (8)$$

For the RTN uplink scenario, the authors in [10] formulate the SINR in a similar way for a specific terminal. However, for the FWD downlink, we formulate the SINR per beam, and all the carriers' SINRs are integrated in an equation (7), the SINR for a specific carrier (8) is also derived from (7).

## IV. FREQUENCY/TIME DUALITY

In the previous section, expression (8) gives the SINR in terms of the spectral mask vector (i.e., in frequency domain). In this section, we will propose the frequency/time duality of (8). For doing so, we first state the dual expression in time domain. After that, we can find the conditions for the duality.

### A. Dual System Model

In the time domain, the time window is segmented into $N_t$ time-slot. The time-slot mask matrix can be formulated as $\mathbf{T} = [\tilde{\mathbf{t}}_1^T, \tilde{\mathbf{t}}_2^T, \cdots, \tilde{\mathbf{t}}_{N_t}^T]^T$, where $\tilde{\mathbf{t}}_j = [t_{1j}, t_{2j}, \cdots, t_{Kj}]$, indicates which beams are allocated time-slot $j$. Then, according to the duality elements in Table I, the transmitted signal power matrix $\mathbf{S}_i^t$, the co-channel interference power matrix $\mathbf{U}_i^t$, the interference power plus the noise matrix $\mathbf{V}_i^t$ and the SINR matrix $\mathbf{\Gamma}_i^t$ in time domain can be formulated as follows

$$\mathbf{S}_i^t = |h_{ii}|^2\mathbf{T}_i\mathbf{T}_i^H, \quad (9)$$

$$\mathbf{U}_i^t = \mathrm{diag}\left\{\left[\tilde{\mathbf{h}}_i\tilde{\mathbf{t}}_j^H\tilde{\mathbf{t}}_j\tilde{\mathbf{h}}_i^H\right]_{j=1,2,\cdots,N_t}\right\}, \quad (10)$$

$$\mathbf{V}_i^t = \mathbf{U}_i^t + \sigma^2 \mathbf{I}_{N_t}, \tag{11}$$

$$\mathbf{\Gamma}_i^t = \mathbf{S}_i^t (\mathbf{V}_i^t)^{-1}. \tag{12}$$

The SINR of beam $i$ and time-slot $j$ will be the $j$th diagonal element of the matrix $\mathbf{\Gamma}_i^t$. Hence, the SINR can be formulated as

$$\gamma_{ij}^t = \frac{|h_{ii} t_{ij}|^2}{\sum_{k=1(k\neq i)}^{K} |h_{ik} t_{kj}|^2 + \sigma^2}. \tag{13}$$

From a theoretical point of view [14], (8) and (13) are dual of each other. However, for a practical system, we derive the duality conditions in the next section.

### B. Duality Conditions

From (8) and (13) we can extract the duality conditions. In order to do so, we first express the beam-level sum-rate throughput as follows

$$R_i^f = \sum_{j=1}^{N_c} \frac{B_{\text{tot}}}{N_c} \eta_{ij}^f, \tag{14}$$

and the dual is

$$R_i^t = \sum_{j=1}^{N_t} \frac{B_{\text{tot}}}{N_t} \eta_{ij}^t, \tag{15}$$

where $\eta_{ij} = f(\gamma_{ij})$ is the spectral efficiency ($\eta_{ij}$ can be $\eta_{ij}^f$ or $\eta_{ij}^t$ in frequency domain or time domain, respectively), and $f(\gamma_{ij})$ equals to $\log_2(1+\gamma_{ij})$ for Shannon limit with Gaussian coding, or can be a quasi-linear function in DVB-S2 [19] with respect to SINR .

Hence, in order to obtain the duality conditions, we assume that $R_i^f = R_i^t$ and the throughput rate per-carrier in frequency domain (or per time-slot in time domain ) is equivalent for each illuminated beam. The following conditions should be fulfilled for systems to be dual in practice:

- Granularity in frequency and time domains should be the same: $N_c = N_t$.
- The entries of resource allocation matrix should be the same in frequency and time domains: $w_{ij} = t_{ij}$.
- The spectral efficiency function $f(\cdot)$ should be the same for NOFR and BH systems in frequency and time domains, respectively: $f(\eta_{ij}^f) = f(\eta_{ij}^t)$.

## V. Capacity Optimization

In this section, we propose two capacity optimization problems under the constraints of the traffic requested per-beam and overall available power. The first problem, P1: capacity optimizing with co-channel interference, maximizes the total capacity allocated with respect to the traffic requested. Since the problem is not only non-convex but also to the need of preserving the geometry of the matrix $\mathbf{W}$. Therefore, we propose an iterative algorithm solution. The second one, P2: capacity optimizing without co-channel interference, is a simplified problem by assuming that the co-channel interference is negligible, hence, the problem can be solved straightforwardly by the lagrangian approach.

TABLE II
Algorithm solution for NOFR system

| |
|---|
| 1: Initialize: $R_k \Leftarrow 0, \forall k$. $n_{\text{it}} \Leftarrow 0$. $\mathbf{W} \Leftarrow \mathbf{0}$ |
| 2: $i \Leftarrow 0$. |
| Generating beam set $\mathcal{A}_s$: |
| $\mathcal{A}_s = \left\{ i_1, i_2, \cdots, i_N \middle\| 0 \leq \frac{R_{i_n}}{\hat{R}_{i_n}} \leq \frac{R_{i_n-1}}{\hat{R}_{i_n-1}} < 1 \right\}$. |
| where $i_n \in \{1, 2, \cdots, K\}, n = 1, 2, \cdots, N$. |
| 3: $n_{\text{it}} \Leftarrow n_{\text{it}} + 1$ |
| **Repeat**: $i \Leftarrow i + 1$. $k \Leftarrow \mathcal{A}_s(i)$ |
| 4: Solve the Rayleigh quotient problem: |
| $\arg\max \frac{\mathbf{e}_j^H \mathbf{S}_k^f \mathbf{e}_j}{\mathbf{e}_j^H \mathbf{V}_k^f \mathbf{e}_j}$ |
| 5: $w_{kj} \Leftarrow \mathbf{e}_j^H \mathbf{e}_j (P_{\text{sat}})^{1/2}$ |
| 6: Update $\mathbf{S}_k^f$, $\mathbf{U}_k^f$. $\mathbf{V}_k^f \Leftarrow \mathbf{U}_k^f + \sigma^2 \mathbf{I}$ |
| 7: go to step 3, |
| **until** $k > i_N$. |
| 8: Update $\gamma_{kj}^f, \forall k, j$. $R_k \Leftarrow \sum_{j=1}^{N_c} \frac{B_{\text{tot}}}{N_c} \eta_{ij}^f, \forall k$. |
| 9: go to step 2, until $\mathcal{A}_s$ is empty or $\sum_{i=1}^{K} \mathbf{w}_i^H \mathbf{w}_i \leq P_{\text{tot}}$. |

### A. P1: capacity optimizing with co-channel interference

Obviously, $\gamma_{ij}^f$ in formula (8) not only depends on the spectral mask vector of beam $i$ ($\mathbf{w}_i$), but also depends on that of the co-channel beams. And hence, the spectral mask vector for each beam must be optimized jointly with the others. The specific design of one beam's spectral mask vector may affect the crosstalk experienced by other beams. Hence it's a complicated task to design the spectral mask matrix $\mathbf{W}$ jointly. In order to best match the offered and the requested traffic per-beam, we develop a methodology to solve the spectral mask matrix $\mathbf{W}$ in this section and to jointly optimize power and carrier allocation. Note that we only discuss the capacity optimization for NOFR system because BH is dual with NOFR, thus the formulation is also applicable for BH system by changing the duality parameters in Table I.

Existing results in the references [20]–[23] is exclusively over the power allocation. However, we assume an additional degree of freedom: carrier allocation (bandwidth granularity). We propose to use binary power allocation (BPA) ($|w_{ij}|^2 = \{0, P_{\text{sat}}\}, i = 1, 2, \cdots, K; j = 1, 2, \cdots, N_c$) and quantized bandwidth allocation in order to decrease the complexity, where $P_{\text{sat}}$ is the TWTA saturation power per carrier.

*1) Optimization Problem Formulation:* The capacity optimization problem can be formulated as

$$\begin{aligned} &\max_{\mathbf{W}} \sum_{i=1}^{K} \frac{R_i(\mathbf{W})}{\hat{R}_i}, \\ &\text{subject to } \; R_i \leq \hat{R}_i, \\ &\sum_{i=1}^{K} \mathbf{w}_i^H \mathbf{w}_i \leq P_{\text{tot}}; \text{and } |w_{ij}|^2 = \{0, P_{\text{sat}}\}, \forall i, j. \end{aligned} \tag{16}$$

where $\hat{R}_i$ is the traffic requested by beam $i$, $R_i(\mathbf{W})$ is defined in Table I. $P_{\text{tot}}$ is total available satellite power, $P_{\text{sat}}$ is the saturation power per carrier, which is limited by the satellite amplifier.

*2) Iterative Algorithm Solution:* The general analytical solution of (16) is a complex problem due not only to the clear non-convexity but also to the need of preserving the geometry of the optimization model (i.e. the structure of matrix $\mathbf{W}$). Therefore, we propose an iterative algorithm solution, which is summarized in Table II. The beam set $\mathcal{A}_s$ is constituted by all the beams, in which the traffic requirement is not achieved (i.e. $\frac{R_k}{\hat{R}_k} < 1$). Quantities associated with the $n$th iteration are denoted by $n_{\text{it}}$. Each iteration is based on a two-step process.

Firstly, we optimize subspace-by-subspace and obtain an analytical solution to the sub-problem of allocating the carrier on a per-beam basis (as shown in step 4 of Table II). The optimal carrier allocation per-beam can be formulated as a generalized Rayleigh quotient, e.g. for beam $i$, the problem can be formulated as:

$$\begin{aligned} &\arg\max_{j} \frac{\mathbf{e}_j^H \mathbf{S}_i^f \mathbf{e}_j}{\mathbf{e}_j^H \mathbf{V}_i^f \mathbf{e}_j}, \\ &\text{subject to} \sum_{i=1}^{K} \mathbf{w}_i^H \mathbf{w}_i \leq P_{\text{tot}}. \end{aligned} \tag{17}$$

where $\mathbf{e}_j \in \mathbb{R}^{N_c \times 1}$ is standard basis vector, which denotes the vector with a "1" in the $j$th coordinate and 0's elsewhere.

The solution of the generalized Rayleigh quotient problem shown in (17) is given as

$$\mathbf{e}_j = \upsilon_{\max}(\mathbf{S}_i^f (\mathbf{V}_i^f)^{-1}) = \upsilon_{\max}(\mathbf{\Gamma}_i^f), \tag{18}$$

where $\upsilon_{\max}(\mathbf{\Gamma}_i^f)$ (as expressed in 7) indicates the eigenvector related to the maximum eigenvalue of matrix $\mathbf{\Gamma}_i^f$.

Secondly, we obtain the power allocated to the selected carriers from the power constraint (as shown in step 5 of Table II). $w_{ij}$ for the $j$th carrier of beam $k$ can be obtained with the solution of $\mathbf{e}_j$ as

$$w_{ij} = \mathbf{e}_j^H \mathbf{e}_j (P_{\text{sat}})^{1/2}. \tag{19}$$

After each iteration, we update matrix $\mathbf{S}_i^f$ and $\mathbf{V}_i^f$ according to the updated spectral mask matrix $\mathbf{W}$.

### *B. P2: capacity optimizing without co-channel interference*

In this section, we assume that the co-channel interference is negligible, because the co-channel beams (in frequency domain) or the simultaneously illuminated beams (in time/space domain) can be separated far from each other in practice. In this way, the capacity optimization can be reduced to a convex problem.

Two cost functions are proposed to solve the frequency and time/space capacity optimizing problem without co-channel interference. Note that we only discuss the optimization problem for BH system because NOFR is dual with BH (see Section IV), thus the formulation is also applicable for NOFR system by changing the related parameters (e.g. $T_s \to B_c$, $N_t \to N_c$).

*1) $n$-th Order Difference Cost Function:* Here we want to match allocated bit rate $R_i$ to requested bit rate $\hat{R}_i$ as closely as possible, i.e., we want to minimize a general function of the difference between $\{R_i\}$ and $\{\hat{R}_i\}$ across all the ground cells.

If an $n$-th order deviation cost function is used, the problem can be formulated as

$$\begin{aligned} \min \quad & \sum_{i=1}^{K} \left| R_i - \hat{R}_i \right|^n, \\ \text{subject to} \quad & R_i, \leq \hat{R}_i \\ & \sum_{i=1}^{K} N_i^t \leq N_{\max}^{\text{re}} N_t. \end{aligned} \tag{20}$$

$N_{\max}^{\text{re}}$ is the number of cells illuminated simultaneously, which is a satellite payload constraint. $N_i^t$ is the number of time-slot allocated to ground cell $i$.

We assume that the power allocated to each time-slot is constant. And hence $R_i$ can be simplified as (assuming Gaussian codes):

$$R_i = \frac{N_i^t}{N_t} B_{\text{tot}} \log_2(1+\gamma_i). \tag{21}$$

Since the co-channel interference is assumed to be negligible, the optimization problem shown in (20) is convex. Therefore, the lagrangian function is given as

$$J(N_i^t) = \sum_{i=1}^{K} \left| R_i - \hat{R}_i \right|^n + \lambda \left( \sum_{i=1}^{K} N_i^t - \right). \tag{22}$$

Let $\frac{\partial J(N_i^t)}{\partial N_i^t} = 0$, we can obtain

$$N_i^t = \frac{\hat{R}_i N_t}{B_{\text{tot}} \log_2(1+\gamma_i)} - \left(\frac{\lambda}{n}\right)^{\frac{1}{n-1}} \left( \frac{N_t}{B_{\text{tot}} \log_2(1+\gamma_i)} \right)^{\frac{n}{n-1}}, \tag{23}$$

where $\lambda$ is the lagrange multiplier and determined from the total available time-slot constraint, which can be obtained by solving the equation

$$\sum_{i=1}^{K} N_i^t \quad = \quad N_{\max}^{\text{re}} N_t. \tag{24}$$

From (23) and (24) we can obtain

$$\lambda = n \left( \frac{\sum_{i=1}^{K} \frac{\hat{R}_i N_t}{B_{\text{tot}} \log_2(1+\gamma_i)} - N_{\max}^{\text{re}} N_t}{\sum_{i=1}^{K} \left( \frac{N_t}{B_{\text{tot}} \log_2(1+\gamma_i)} \right)^{\frac{n}{n-1}}} \right)^{n-1}. \tag{25}$$

If we replace $\lambda$ in (23) with (25), the solution will be

$$N_i^t = \frac{\hat{R}_i N_t}{B_{\text{tot}} \log_2(1+\gamma_i)} - \frac{\sum_{k=1}^{K} \left( \frac{\hat{R}_k N_t}{B_{\text{tot}} \log_2(1+\gamma_k)} \right) - N_{\max}^{\text{re}} N_t}{\sum_{k=1}^{K} \left( \frac{\log_2(1+\gamma_i)}{\log_2(1+\gamma_k)} \right)^{\frac{n}{n-1}}}. \tag{26}$$

With the the number of time-slot allocated to each ground cell ($N_i^t$), the throughput allocated to each cell ($R_i$) can be calculated with (21). We should note that the solution in (26) is independent of the order $n$ ($n \geq 2$) in our case, since we suppose BPA, no co-channel interference and the same channel attenuation factor ($\alpha_i$) for all the ground cell.

*2) Fairness Cost Function:* Another way to match allocated capacity $R_i$ to requested capacity $\hat{R}_i$ is to maximize the ratio between them as

$$\begin{aligned} \max \quad & \prod_{i=1}^{K} \left(\frac{R_i}{\hat{R}_i}\right)^{\omega_i}, \\ \text{subject to} \quad & R_i \leq \hat{R}_i, \\ & \sum_{i=1}^{K} N_i^t \leq N_{\max}^{\text{re}} N_t, \end{aligned} \tag{27}$$

where $\omega_i$ is the weighting factor that represents the priority of each beam. The problem (27) can be easily converted to a convex problem by introducing the logarithm in the objective function. Thus, the optimization problem is converted to

$$\max \quad \sum_{i=1}^{K} \omega_i \log_2 \left(\frac{R_i}{\hat{R}_i}\right). \tag{28}$$

Thus, the lagrangian function is given as

$$J(N_i^t) = -\sum_{i=1}^{K} \omega_i \log_2 \left(\frac{R_i}{\hat{R}_i}\right) + \lambda \left(\sum_{i=1}^{K} N_i^t - N_{\max}^{\text{re}} N_t\right). \tag{29}$$

Let $\frac{\partial J(N_i^t)}{\partial N_i^t} = 0$, then

$$N_i^t = \frac{\omega_i \hat{R}_i N_t}{\lambda \ln 2 B_{\text{tot}} \log_2(1+\gamma_i)}. \tag{30}$$

With given constraint $\sum_{i=1}^{K} N_i^t = N_{\max}^{\text{re}} N_t$, the lagrange multiplier can be solved as

$$\lambda = \frac{\sum_{i=1}^{K} \frac{\omega_i \hat{R}_i N_t}{B_{\text{tot}} \log_2(1+\gamma_i)}}{N_{\max}^{\text{re}} N_t \ln 2}. \tag{31}$$

The solution will be (replace $\lambda$ in (30) with (31))

$$N_i^t = \frac{\omega_i \hat{R}_i N_t}{\log_2(1+\gamma_i)} \frac{N_{\max}^{\text{re}} N_t}{\sum_{k=1}^{K} \frac{\omega_k \hat{R}_k N_t}{\log_2(1+\gamma_k)}}. \tag{32}$$

Therefore, the throughput allocated to each ground cell ($R_i$) can be calculated with (21).

## VI. Technological Gap

From Section IV-B we can see that the spectral efficiency that each technology can provide makes the real difference. Therefore, NOFR and BH systems are not completely dual of each other. In this section, we will demonstrate the technological gap between NOFR and BH. Note that we only consider the forward (FWD) downlink, because the FWD uplink is not a big issue since power at the gateway can be greatly increased to compensate the attenuation.

Equivalent isotropically radiated power (EIRP) is defined as (in dB)

$$EIRP = P_{\text{sat}} - OBO - L_{\text{repeater}} - L_{\text{antenna}} + G_{\text{tx}}, \tag{33}$$

where Output BackOff (OBO) is the ratio of maximum output (saturation) power to actual output power, $L_{\text{repeater}}$ is the repeater loss, $L_{\text{antenna}}$ is the antenna feed loss, and $G_{\text{tx}}$ is the satellite Tx. antenna gain. With known EIRP, we can obtain FWD downlink $C/N_0$ (in dBHz) and SNR (in dB) as

$$C/N_0 = EIRP - L_{\text{propagation}} + (G/T)_{\text{gt}} - 10\log_{10}(k_B), \tag{34}$$

$$SNR = C/N_0 - 10\log_{10}(B_c), \tag{35}$$

where $L_{\text{propagation}}$ is the propagation loss, $(G/T)_{\text{gt}}$ is the ground terminal $G/T$ and $k_B$ is the Boltzmann constant. Let $a = P_{\text{sat}} - L_{\text{repeater}} - L_{\text{antenna}} + G_{\text{tx}} - L_{\text{propagation}} + (G/T)_{\text{gt}} - 10\log_{10}(k_B) - 10\log_{10}(B_c)$, and let $x_1$ and $x_2$ be the OBO for NOFR and BH systems, respectively. Therefore, (35) can be reformulated as

$$SNR^f = a - x_1, \text{ or } SNR^t = a - x_2. \tag{36}$$

Let the FWD downlink signal to co-channel interference SIR be given as $y$, Therefore, the FWD downlink SINR can be formulated as

$$SINR_{\text{down}}^{-1} = SIR^{-1} + SNR^{-1} = y^{-1} + 10^{-(\frac{a-x}{10})}, \tag{37}$$

where $x$ can be $x_1$ or $x_2$ and $SNR$ can be $SNR^f$ or $SNR^t$ for NOFR or BH system.

Let the FWD uplink SINR be $z$, then the FWD whole link SINR is given as

$$SINR_{\text{tot}}^{-1} = SINR_{\text{up}}^{-1} + SINR_{\text{down}}^{-1} = z^{-1} + y^{-1} + 10^{\frac{x-a}{10}}. \tag{38}$$

Let the whole FWD link SINR be $\gamma = SINR_{\text{tot}}$, the spectral efficiency in the case of Shannon limit with Gaussian coding can be given as

$$\eta = \log_2(1+\gamma) \simeq \log_2(\gamma) = -\log_2(z^{-1} + y^{-1} + 10^{\frac{x-a}{10}}), \tag{39}$$

where we make a high SINR approx given as, $\log_2(1+\gamma) \simeq \log_2(\gamma)$. Therefore, the spectral efficiency for NOFR and BH system are

$$\eta^f = -\log_2(z^{-1} + y^{-1} + 10^{\frac{x_1-a}{10}}), \tag{40}$$

$$\eta^t = -\log_2(z^{-1} + y^{-1} + 10^{\frac{x_2-a}{10}}). \tag{41}$$

Let the technological gap of spectral efficiency between BH and NOFR system $\Delta\eta$ be given as

$$\Delta\eta = \eta^t - \eta^f = \log_2 \frac{z^{-1} + y^{-1} + 10^{\frac{x_1-a}{10}}}{z^{-1} + y^{-1} + 10^{\frac{x_2-a}{10}}}. \tag{42}$$

Let $z$, $x_1$ and $x_2$ be constant and $x_1 > x_2$, $\Delta\eta$ will be a monotonically increasing function of $y$. Therefore, the upper bound (maximum) of the technological gap $\Delta\eta$ will be

$$\Delta\eta_{\max} = \Delta\eta|_{y\to+\infty} = \log_2 \frac{1 + z10^{-(\frac{a-x_1}{10})}}{1 + z10^{-(\frac{a-x_2}{10})}}. \tag{43}$$

As we indicated before, the uplink is not relevant. Thus we can suppose that the uplink SINR $z$ is constant. The result of the technological gap is demonstrated in Fig.8, it is meaningful for us to evaluate the performance of NOFR and BH, and to predict the technological gap between NOFR and BH systems.

## VII. Numerical Results

The objective of the simulation is: Firstly, to evaluate the performance of the proposed capacity optimization approaches. Secondly, to compare the proposed system design with the conventional design, which is regular frequency reuse ($f_R = 7$) and uniform power allocation. Thirdly, to obtain the technological gap for a realistic implementation. The payload parameters are defined in [24].

In order to fairly compare the performance with different number of beams in the same coverage (e.g. the European countries), we assume that the total traffic requirement is the same for all the cases. The linear traffic distribution model is defined as $\hat{R}_k = k\beta; k = 1, 2, \cdots, K$, $\beta$ is slope of the linear function. The following parameters are assumed in the simulations: $P_{\text{sat}} = 4$Watt, $B_{\text{tot}} = 500$MHz, $N_c = 112$, each cluster contains 7 beams (as shown in Fig.2), $\beta = 8 \times 10^6$bps for $K = 121$, and $B_c = B_{\text{tot}}/N_c = 4.4643$MHz.

The parameters of power gain ($g$), spectral efficiency ($\eta$) and traffic matching ratio ($\rho$) are studied in the simulation, which are defined as the following.

### A. Performance Parameters Definition

*1) Power Gain:* We compare the amount of total power consumption for joint power and bandwidth optimized allocation with that for uniform power and bandwidth allocation when both achieve the same useful throughput using the same total bandwidth. We define the power gain $g_p$ as

$$g_p = \frac{K P_{\text{uni}}}{\sum_{k=1}^{K} \mathbf{w}_k^H \mathbf{w}_k}, \tag{44}$$

where $P_{\text{uni}}$ denotes the power per-beam of the uniform allocation scheme.

*2) Spectral Efficiency:* The spectral efficiency is defined based on the total allocated traffic and total allocated bandwidth as

$$\eta = \frac{\sum_{k=1}^{K} R_k}{\sum_{k=1}^{K} B_k}. \tag{45}$$

*3) Traffic Matching Ratio:* In order to describe the satisfaction degree of the allocated traffic with respect to the total request traffic, the traffic matching ratio is defined here as

$$\rho = \frac{\sum_{k=1}^{K} R_k}{\sum_{k=1}^{K} \hat{R}_k}. \tag{46}$$

### B. Beam Layout and Antenna Model

We assume a general beam layout model (shown in Fig.2). A fixed-size space is used to generate different number of beams, thus, the beamwidth is decreasing as the number of beams increases. It means that the larger the number of beams, the narrower the beamwidth. We assume a tapered aperture antenna model with 47.14 dBi maximum antenna gain. Then the SINR can be calculated in each iteration of the algorithm with a given link budget of a typical Ka-Band (19.95 GHz) satellite payload.

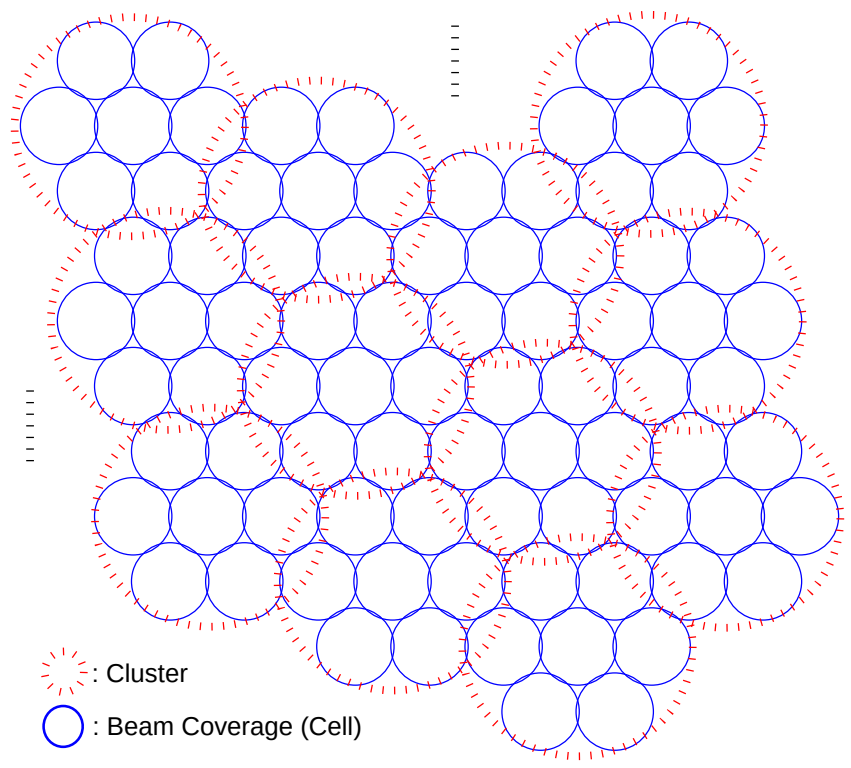

Fig. 2. Beam layout.

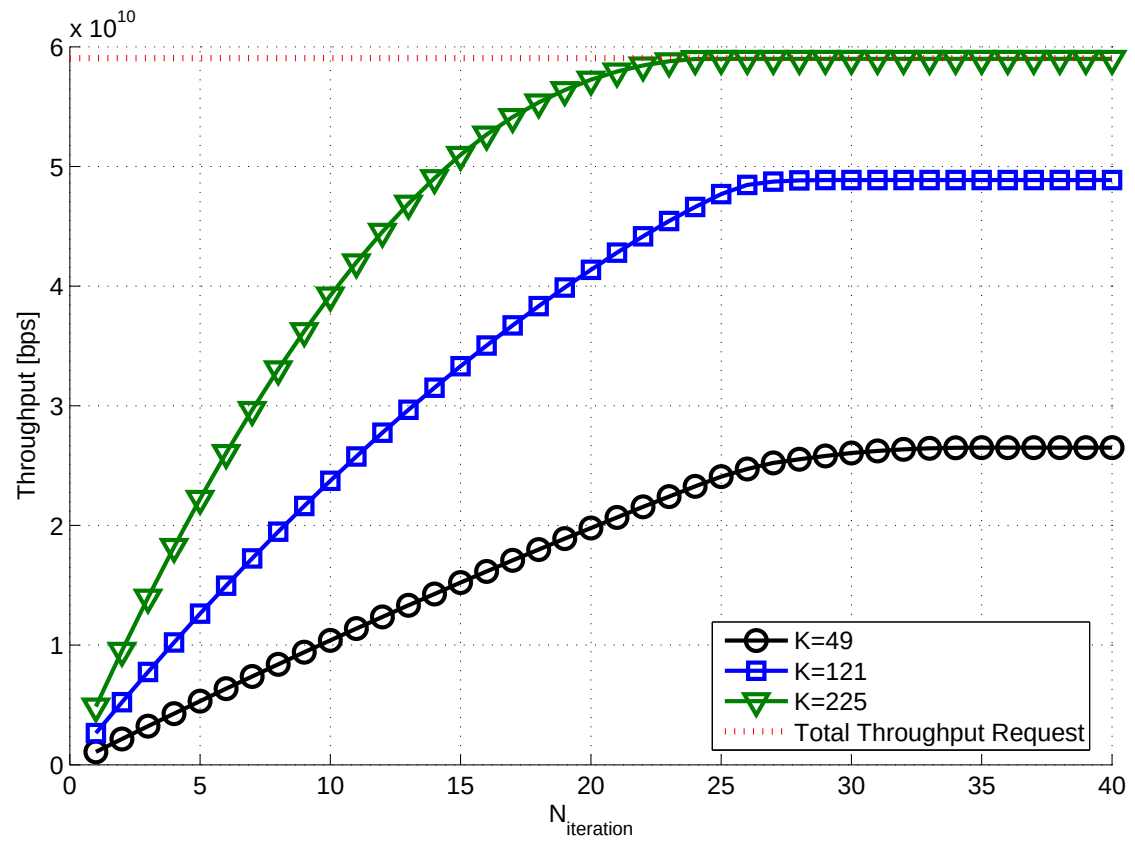

Fig. 3. Convergence speed.

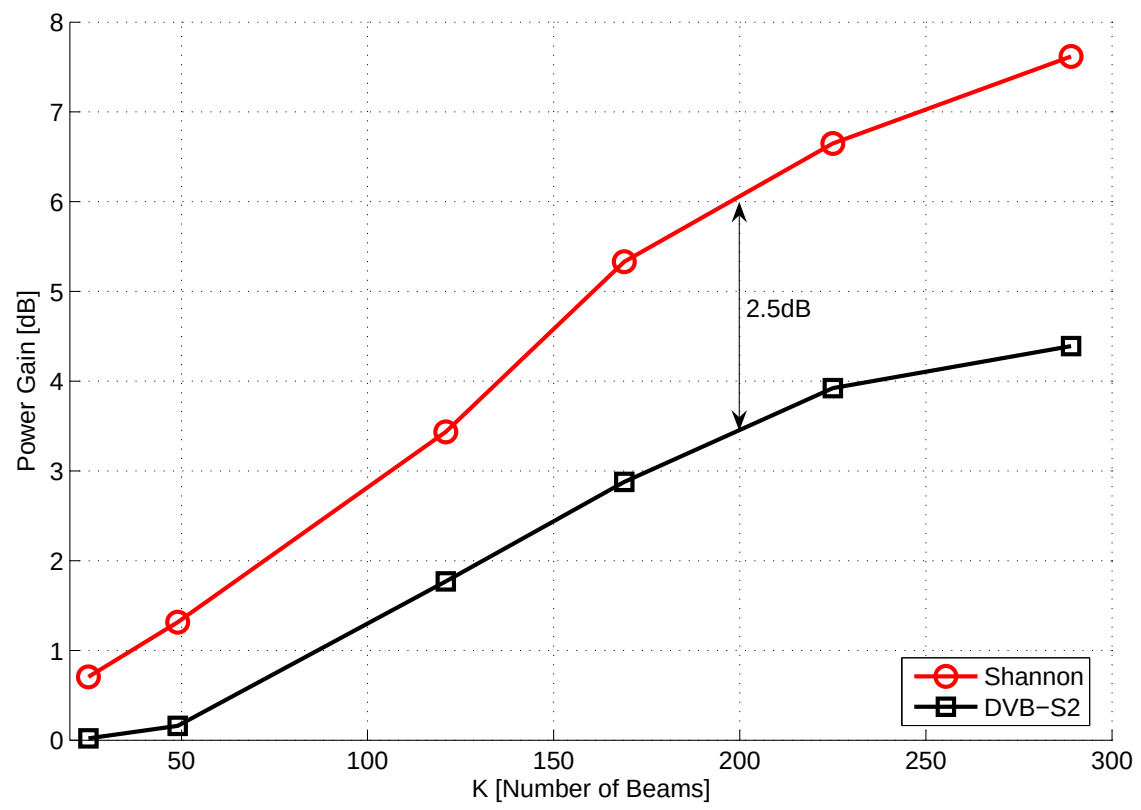

Fig. 4. Power gain ($g_p$) vs. number of beams ($K$).

### C. Simulation Results

In order to evaluate the relevance of our iterative algorithm (shown in Table II), we perform a study of convergence. It can be observed from Fig.3 that the algorithm is convergent for different number of beams, and the convergence is faster with the number of beams increasing, e.g. our algorithm runs 24 and 33 iterations for number of 225 and 49 beams respectively.

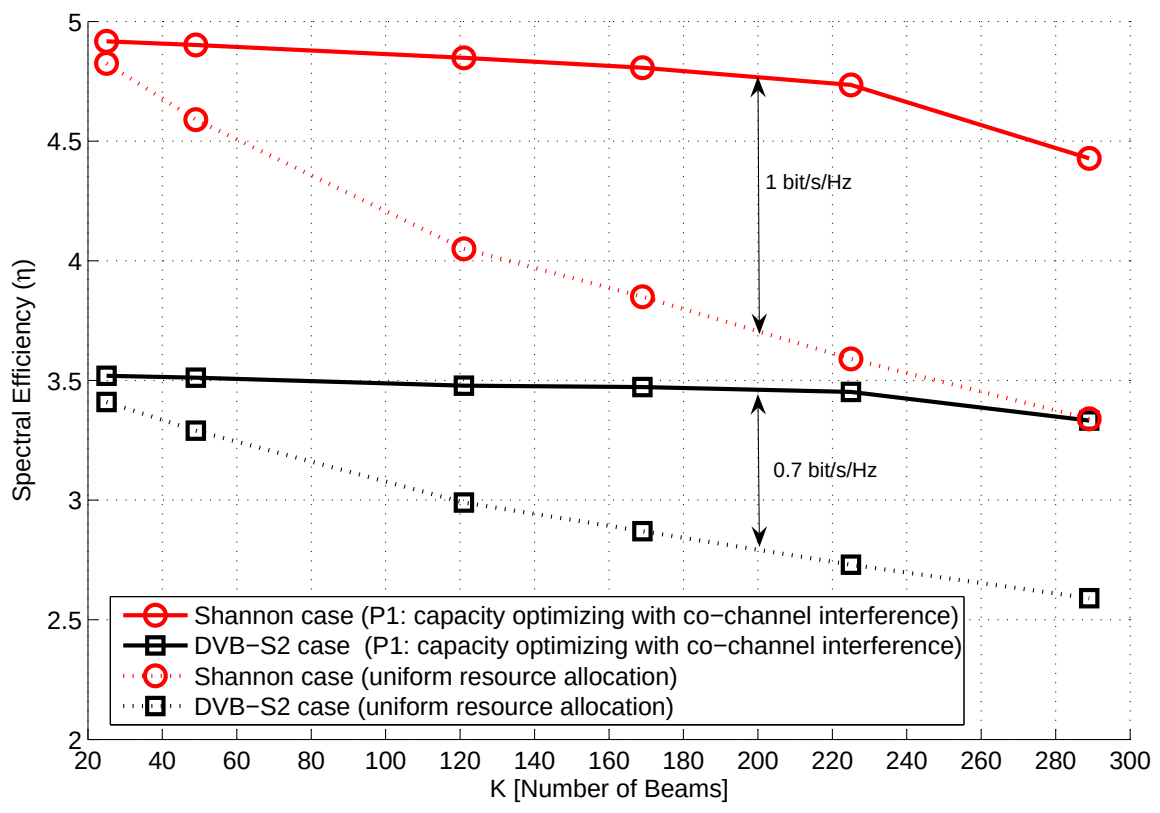

Fig. 5. Spectral efficiency ($\eta$) vs. number of beams ($K$).

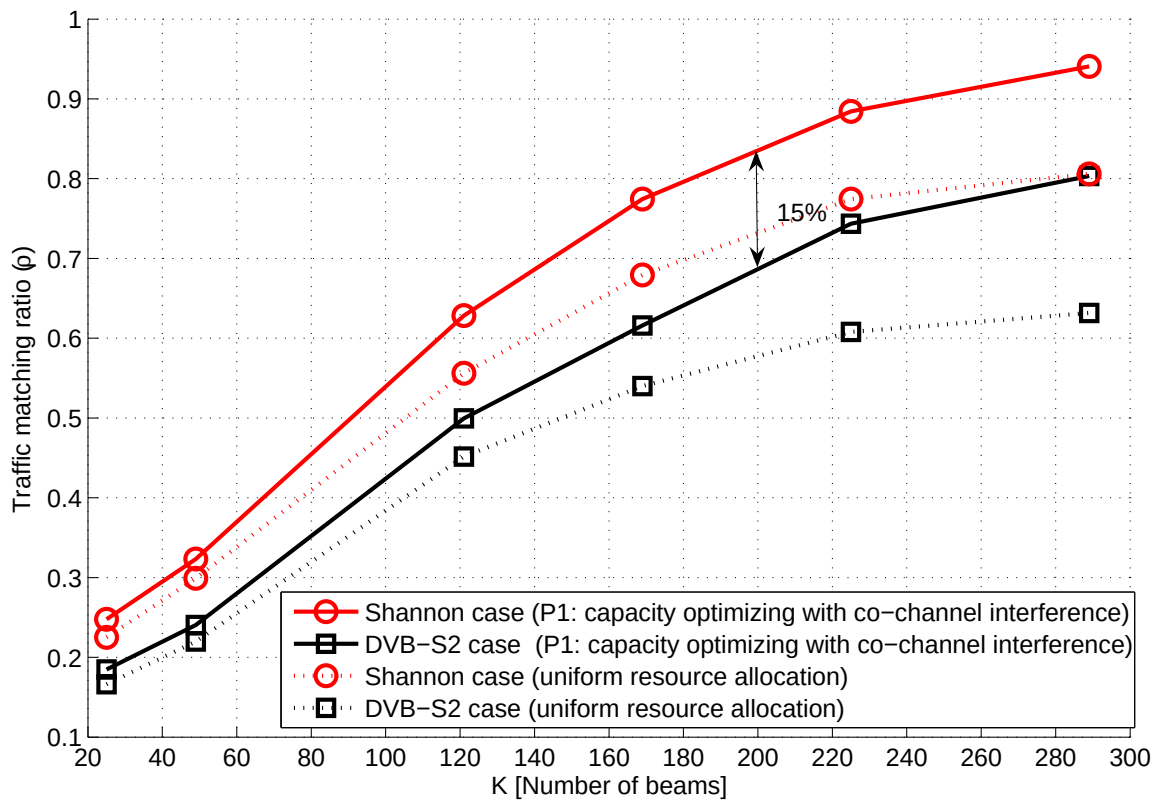

Fig. 6. Traffic matching ration ($\rho$) vs. number of beams ($K$).

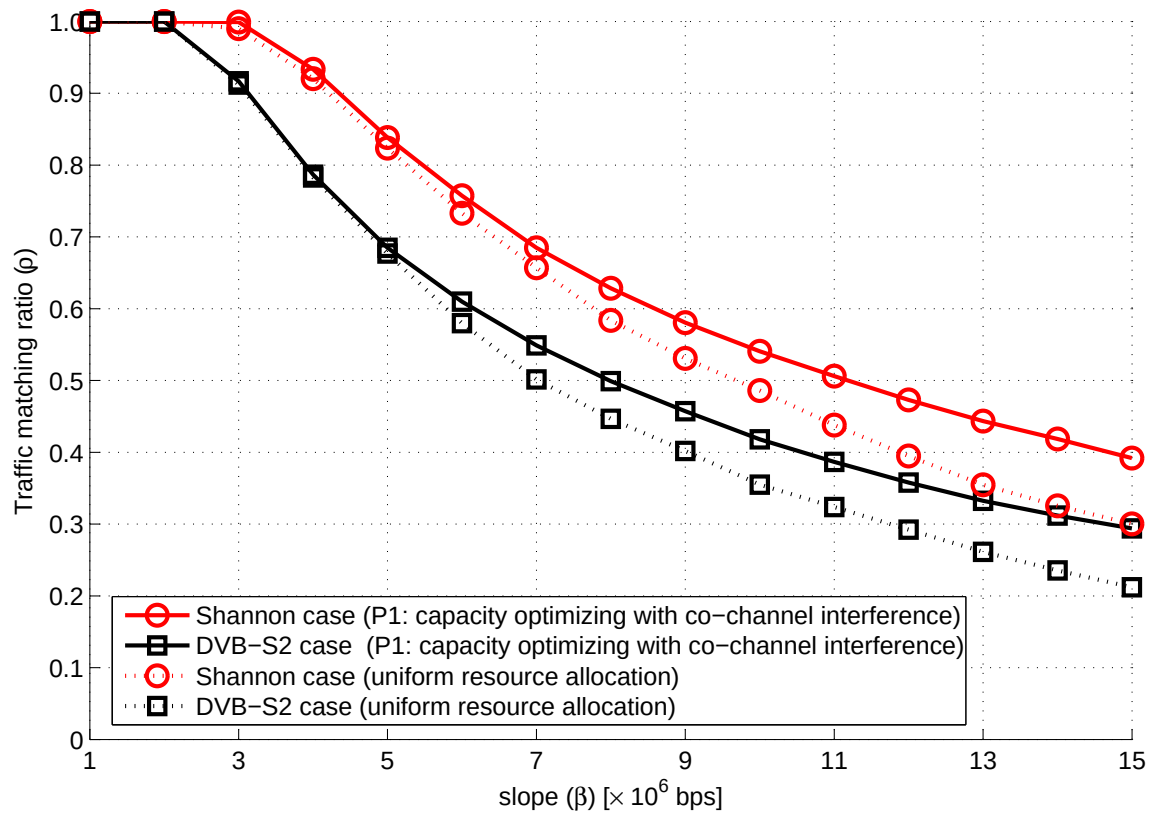

Fig. 7. Traffic matching ration ($\rho$) vs. traffic slope ($\beta$).

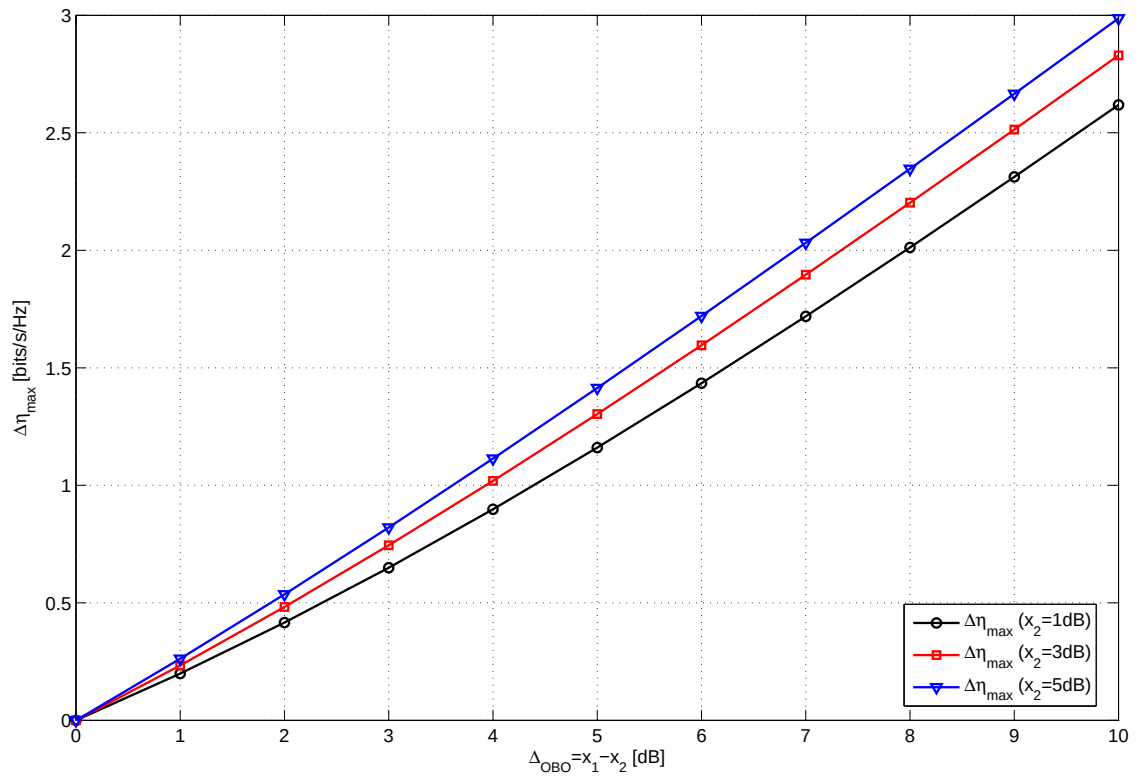

Fig. 8. $\Delta_{\text{OBO}}$ vs. $\Delta\eta_{\max}$.

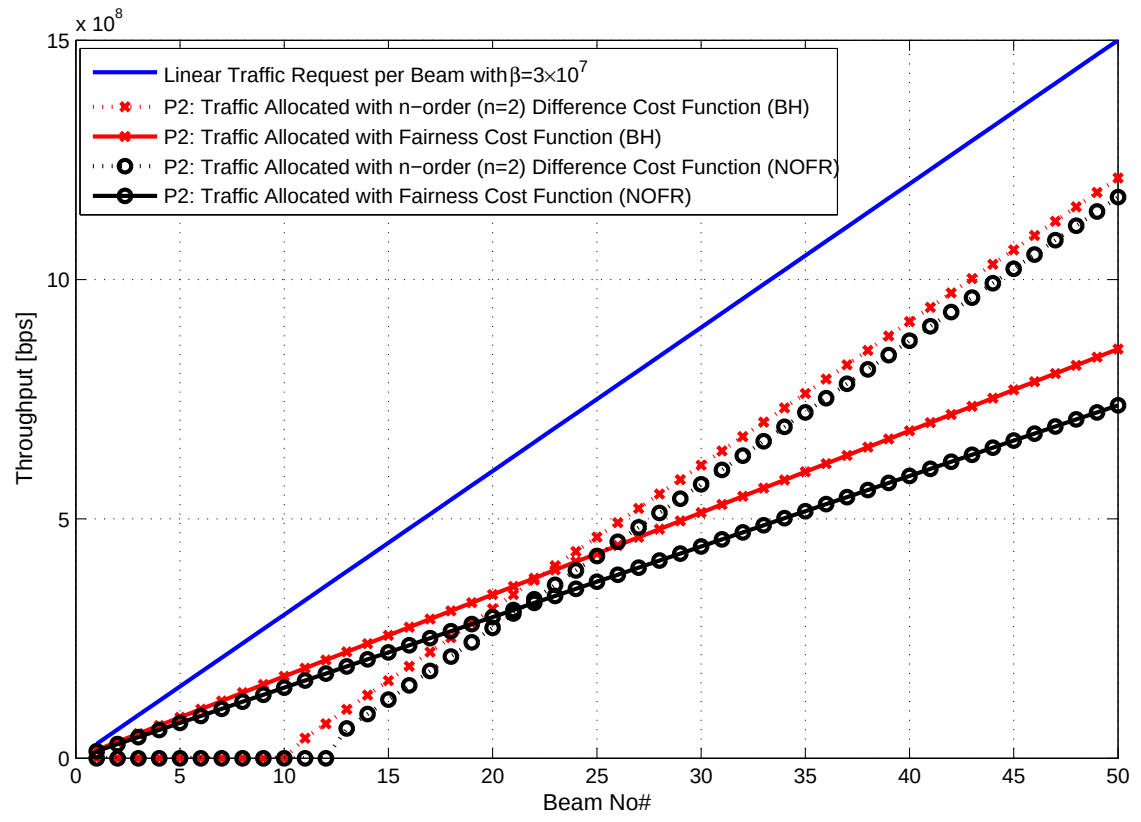

Fig. 9. Comparison of cost functions in terms of throughput.

The reason is that the algorithm allocates resources to all unsatisfied beams in each iteration, thus, more traffic will be allocated with larger number of beams. Consequently, all the beams will reach the traffic requirement faster. The algorithm has been applied in the realistic satellite payload model and proved in [24] that it is applicable to the current multibeam satellite system.

The power gain with respect to the number of beams is shown in Fig.4. We can see that about 6dB and 3.5dB power gain can be achieved by capacity optimizing of P1 with Gaussian coding and DVB-S2 ModCods (defined in [19]), respectively (when $K = 200$). By optimizing the capacity achieved per-beam, we do not only reduce the power and bandwidth consumption of small traffic requirement beams, but also achieve reasonable proportional fairness from the viewpoint of user beams. In Fig.5, the result shows that the spectral efficiency decreases with the number of beams increasing, especially when $K > 200$. The reason is that co-channel interference will increase with the beamwidth decreasing.

In Fig.6 we can observe that the traffic matches better in case of larger number of beams. However, the power consumption and the complexity will increase with larger number of beams. Therefore, we should balance the total achieved throughput with respect to both power consumption and complexity. Fig.7 shows the traffic matching ratio with respect to different traffic distribution slope. Obviously, the traffic matching ratio drops down with the slope increasing. Because the traffic distribution is more asymmetric with larger slope. We can see that the capacity optimization can achieve better matching ratio compared to the conventional design for both Shannon and DVB-S2 cases.

In order to evaluate the technology gap, we define the difference of OBO between NOFR and BH systems as $\Delta_{\text{OBO}} = x_1 - x_2$. Fig.8 shows $\Delta_{\text{OBO}}$ with respect to $\Delta\eta_{\max}$, which is defined in (43). We can see that $\Delta\eta_{\max}$ is almost linear with $\Delta_{\text{OBO}}$, and the slope is increasing with BH system OBO ($x_2$) increasing. This result is very useful to predict the technological gap between NOFR and BH systems.

Fig.9 shows the distribution of throughput for $n$-order difference cost function and fairness cost function along $K = 50$

beams that have a linear distribution traffic demand. In this simulation, we assume that $\beta = 3 \times 10^7$, $n = 2$ (second order function), $N_t = 32$, $N^{\text{re}}_{\text{max}} = 8$, the SINR $\gamma_k$ and the weighting factor $\omega_k$ are constant for all the cells in order to simplify. The result shows that two different cost functions distribute the total available resource (carriers or time/space) to all the ground cells with different pattern. Fairness cost function is more favorable for low traffic requirement cells while $n$-order cost function distribute more resource to high traffic requirement beam. The performance of BH is slightly better than NOFR, especially for the low traffic requirement beams. Further, the $n$-order simply neglect too low-loaded beams. This is relevant result since it is already considered in satcom design.

## VIII. Conclusions

Current designs of broadband satellite systems are lack of the necessary flexibility to match realistic asymmetric traffic distributions. Two new technologies are studied to replace the current ones over multibeam satellite systems. We prove that the two technologies are dual of each other in frequency and time domains. Moreover, the technological gap between NOFR and BH systems is formulated. Two novel capacity optimization problems, P1 and P2, are investigated to best match the individual SINR constraints. The current state-of-the art PHY layer technology: DVB-S2 and Shannon are implemented in order to evaluate the performance. The results show significant improvements in terms of power gain, spectral efficiency and traffic matching ratio compared to the conventional system. For a DVB-S2 and $K = 200$ case, we can achieve up to 3 dB power gain, 0.7 bit/s/Hz spectral efficiency gain, and improve $10\%$ traffic matching ratio by the proposed capacity optimizing approach. For the duality study, the results show that the technological gap is only related to the OBO of NOFR and BH, and the gap is almost linear with $\Delta_{\text{OBO}}$. Further, we solve the second problem P2 with different cost functions. Fairness cost function is more favorable for low traffic requirement cells while $n$-order cost function distribute more resource to high traffic requirement beams. The study of the capacity optimization shows that the BH system performs only slightly better than NOFR. We also prove the primary goal of the study, that the NOFR and BH technologies can match much better than the conventional design in the asymmetrical traffic distribution model.

## Acknowledgment

The authors would like to acknowledge the support of European Space Agency (ESA) under the project of "Beam Hopping techniques for multibeam satellite systems".